\documentclass[12pt,reqno]{amsart}
\pdfoutput=1 
\usepackage{amsmath,bbm}
\usepackage{latexsym}
\usepackage{amsfonts}
\usepackage{amssymb}
\usepackage{color}
\usepackage{graphicx}
\usepackage{url}
\usepackage{enumerate}
\usepackage{tikz}
\usetikzlibrary{shadings, intersections, calc, plotmarks}
\usepackage{marginnote}
\usepackage{stackrel}
\usepackage{mathtools}
\usepackage[colorlinks=true,linkcolor=blue,citecolor=blue,urlcolor=blue,plainpages=false,pdfpagelabels]{hyperref}

\usepackage{geometry}
\xdefinecolor{tumblue}     {RGB}{0,101,189}
\xdefinecolor{tumgreen}    {RGB}{162,173,  0}
\xdefinecolor{tumred}      {RGB}{229, 52, 24}
\xdefinecolor{tumivory}    {RGB}{218,215,203}
\xdefinecolor{tumorange}   {RGB}{227,114, 34}
\xdefinecolor{tumlightblue}{RGB}{152,198,234}

\newtheorem{theorem}{Theorem}
\newtheorem*{theorem*}{Theorem}
\newtheorem{lemma}{Lemma}

\newtheorem*{corollary*}{Corollary}

\newtheorem{remark}{Remark}

\newtheorem{definition}{Definition}

\newcommand{\R}{\mathbbm{R}}
\newcommand{\C}{\mathbbm{C}}
\newcommand{\N}{\mathbbm{N}}

\newcommand{\F}{\mathbbm{F}}
\newcommand{\Q}{\mathbbm{Q}}

\newcommand{\1}{\mathbbm{1}}

\def\>{{\rangle}}
\def\<{{\langle}}

\newcommand{\be}{\begin{equation}}
\newcommand{\ee}{\end{equation}}
\newcommand{\bea}{\begin{eqnarray}}
\newcommand{\eea}{\end{eqnarray}}

\newcommand{\ket}[1]{|#1\rangle} 
\newcommand{\bra}[1]{\langle#1|} 
\newcommand{\tr}[1]{\mathrm{tr}\left[#1\right]} 
\newcommand{\Tr}{\mathrm{tr}}

\newcommand{\margintxt}[1]{}

\begin{document}

\title[Transcendental properties of entropy-constrained sets: Part II]{Transcendental properties of entropy-constrained sets: Part II}

\author[Blakaj]{Vjosa Blakaj$^{1,2}$}
\email{vjosa.blakaj@tum.de}
\author[Manai]{Chokri Manai$^{1,2}$}
\email{chokri.manai@tum.de}
\address{$^1$ Department of Mathematics, Technical University of Munich, Garching, Germany}
\address{$^2$ Munich Center for Quantum
Science and Technology (MCQST),  M\"unchen, Germany}

\begin{abstract} 
In this work, we address the question of the impossibility of certain single-letter formulas by exploiting the semi-algebraic nature of various entropy-constrained sets. The focus lies on studying the properties of the level sets of relative entropy, mutual information, and R\'{e}nyi entropies. We analyze the transcendental structure of the set of states in which one of the aforementioned entropy quantities is fixed. Our results rule out (semi)algebraic single-shot characterizations of these entropy measures with bounded ancilla for both the classical and quantum cases.\vspace{10pt}

\end{abstract}

\maketitle
\tableofcontents

\section{Introduction}\label{sec:intro}
Algebraic geometry has been a useful tool in the study of various problems in quantum information theory \cite{FCS, WolfCubbitPerez, Vjosa, MagicSquares}. The distinction between the semialgebraic world and the transcendental world has proven useful in separating the single-shot and asymptotic settings \cite{Vjosa}. Finite resource theories are often described by polynomial (in)equalities since large parts of the underlying theory are described by relatively small, simple mathematical objects defined on finite-dimensional vector spaces. Infinite resources, on the other hand, fall outside this regime and could be associated with the transcendental world, and the difficulty of studying such regimes arises from quantification over infinite dimensional underlying structures \cite{WolfCubbitPerez}.\vspace{5pt}

Our main motivation for this work lies in the question of the impossibility of single-letter formulas, especially for asymptotically defined quantities. We study this by exploiting the transcendental properties of certain entropy-constrained sets. The results we provide here are based on several characterizations of algebraic functions and on the fact that von Neumann's entropy-constrained sets are nowhere semialgebraic. The latter was proved in \cite{Vjosa} exploiting the fact that the analytic continuation of algebraic functions has at most a finite number of branches.  \vspace{5pt}

There have been many attempts to determine whether the entropy quantities that usually characterize the asymptotic regime can be given operational meaning in the single-shot setting \cite{CatalyticEntropy, RevCatalysis, Conjecture,RelativeEntropy, PureEntangled}. One of the many interesting problems that could be investigated from a (semi)algebraic point of view is that of entanglement catalysis. These catalytic state transformations have been studied in \cite{PureEntangled, EntanglementCatalysis} and are defined as follows: \vspace{3pt}

For a given bipartite entangled state shared between two parties, say, Alice and Bob, $\ket{\psi}\bra{\psi}^{AB}$ on $\C^{m_A} \otimes \C^{m_B}$, define $\mathcal{C}_{m,d}$ as the set of all entangled states $\ket{\phi} \bra{\phi}^{AB}$ on $\C^{m_A} \otimes \C^{m_B}$ such that for any $\varepsilon > 0$ there is a state $\tau^{A'B'}$ on $\C^{d_{A'}} \otimes \C^{d_{B'}}$ and a LOCC protocol $\Lambda$ satisfying 
\begin{align}
    \Tr_{AB}[\Lambda(\ket{\psi}\bra{\psi}^{AB} \otimes \tau^{A'B'})] & = \tau^{A'B'}, \\
    |\!| \ket{\phi}\bra{\phi}^{AB} - \Tr_{A'B'}[\Lambda(\ket{\psi}\bra{\psi}^{AB} \otimes \tau^{A'B'})] |\!|_1 & \leq \varepsilon, \\
    |\!| \Lambda(\ket{\psi}\bra{\psi}^{AB} \otimes \tau^{A'B'}) - \ket{\phi}\bra{\phi}^{AB} \otimes \tau^{A'B'} |\!|_1 & \leq \varepsilon.
\end{align}

Such an LOCC protocol $\Lambda$ can be described by the following three steps as a consequence of Theorem $1$ in \cite{PureEntangled}: as a first step, Alice performs rank $1$ projective measurement on her auxiliary system and depending on the outcome, the other parties apply a certain LOCC protocol $\Gamma$, or not. Then Alice continues with an application of a unitary on the auxiliary system, and as a last step, all parties perform a SWAP unitary. By the results in \cite{OneLOCC} the LOCC protocol $\Gamma$ is in fact equivalent to a strategy involving \textit{only} a single (generalized) measurement by Alice, followed by a one-way communication of the result to Bob. In other words, $\mathcal{C}_{m,d}$ is the set of states that can be reached (approximately) from $\ket{\psi}\bra{\psi}^{AB}$ using a bounded catalyst $\tau^{A'B'}$, and everything defined on this set comes from a bounded vector space and can be written in terms of polynomial (in)equalities. 

As a second set, consider the set $\mathcal{C}_m$ of all pure states on $\C^{m_A} \otimes \C^{m_B}$ whose entanglement entropy is smaller than or equal to that of the initial state $\ket{\psi}\bra{\psi}^{AB}$. If no bound is assumed on the dimension of the catalyst, then it is known that $\mathcal{C}_{m,\infty} = \mathcal{C}_m$ \cite{PureEntangled, EntanglementCatalysis}. The question now arises whether there is equality between these two sets for some $d$ as a function of $m$. One way to settle this question and rule out equality is to state that due to the Tarski-Seidenberg theorem ($1.4$ and $2.2$, \cite{RAG98}), the set $\mathcal{C}_{m,d}$ is semialgebraic, while the set $\mathcal{C}_m$ is not, as shown in \cite{Vjosa}. This observation shows that there is no universal bound on the dimension of the catalyst for the catalytic LOCC transformations presented in \cite{PureEntangled}. Even more, no semialgebraic characterization of the entropy-constrained sets would be possible as long as a bounded ancilla is considered.

\vspace{10pt}

\begin{table}[h!]\label{Choko}
\setlength{\arrayrulewidth}{0.1mm}
\setlength{\tabcolsep}{12pt}
\renewcommand{\arraystretch}{1.5}
\begin{tabular}{ |p{5cm}|p{3cm}|p{3cm}|  }
\hline
\multicolumn{3}{|c|}{Algebraic nature of entropy measures} \\
\hline
Function & Space dimension &  Level set \\
\hline
von Neumann entropy & $d = 2$ & (semi)algebraic everywhere \\
\hline
von Neumann entropy & $d \geq 3$ & transcendental everywhere \\
\hline
Relative entropy $S(\rho || \sigma)$ 

with $\sigma$ fixed & $d \geq 3$   &  transcendental everywhere \\
\hline
Relative entropy $S(\rho || \sigma)$ 

with $\rho$ fixed & $d \geq 3$  & transcendental everywhere \\
\hline
Relative entropy

$S(\rho || \sigma)$   & $d \geq 3$ & transcendental everywhere \\
\hline
Mutual information

$I(\rho_{AB}) \coloneqq  S(\rho_{AB}||\rho_A \otimes \rho_B)$ & $\min(d_A, d_B)\geq 3$ & transcendental \\
\hline
R\'{e}nyi entropy in the limit

$\alpha \to 0$ & $d \geq 2$ & (semi)algebraic everywhere   \\
\hline
R\'{e}nyi entropy in the limit 

$\alpha \to \infty$  & $d \geq 2$ & (semi)algebraic everywhere  \\
\hline
R\'{e}nyi entropy with

$\alpha \in \Q \cap [(0,1) \cup (1, \infty)]$  & $d \geq 2$ & (semi)algebraic everywhere  \\
\hline
R\'{e}nyi entropy with

$\alpha \in (\R \setminus \Q) \cap [(0,1) \cup (1, \infty)]$  & $d = 2$ & (semi)algebraic everywhere \\
\hline

R\'{e}nyi entropy with 

$\alpha \in (\R \setminus \Q) \cap [(0,1) \cup (1, \infty)]$  & $d \geq 3$ & transcendental everywhere \\
\hline

\end{tabular}
\vspace{10pt}
\caption{Overview of the algebraic behavior of entropy level sets.}
\end{table}

\emph{Outline of the paper.} 
After establishing the necessary notations for the paper in Section \ref{sec:prelim}, and specifying the proof method on the following subsection, we give the complete proof of the surfaces of the relative entropy in Section \ref{Sec:RelativeEntropy}. This is divided into three parts, with all variables considered accordingly. Following the same train of thought, we analyze the nature of $\alpha$-R\'{e}nyi entropy constrained sets in Section \ref{sec:Renyi} and show that these sets are nowhere semialgebraic when parameterized by an irrational number. On the other hand, when this parameter is a rational number, the $\alpha$-R\'{e}nyi entropy level sets are everywhere semialgebraic. For both proofs, the classical and limiting cases are discussed in parallel with the quantum case. The argument in Section \ref{Sec:MutualInfo} provides a global result for the level sets of mutual information. We give an overview of the (semi)algebraic nature of the level sets of entropy measures studied here and in \cite{Vjosa}, without considering their extremal values, in Table \ref{Choko}.\vspace{5pt}

\emph{Proof method.}
To establish transcendental properties of entropy-constrained sets, we mainly invoke ideas and techniques from algebraic geometry and complex analysis. For instance, the well-known Tarski-Seidenberg theorem is used several times to derive the semialgebraic nature of various sets. The result of \cite{Vjosa} on the transcendence of von Neumann's entropy-constrained sets is of particular importance for the study of the level set of mutual information. However, the proofs on $\alpha-$R\'{e}nyi entropies and relative entropy require substantially new ideas. In these cases, on the one hand, we take advantage of the nature of algebraic singularities, which precludes the occurrence of poles of irrational degrees. On the other hand, a careful and detailed analysis invoking several ideas from complex analysis  allows us to establish irrational poles of the local level function of the constrained sets.

\section{Notation}\label{sec:prelim}

A set $S \subseteq \R^n$ is called \textit{semialgebraic} if it is defined by a finite number of polynomial equations and inequalities; otherwise, the set is called \textit{transcendental}. Unless differently stated, all polynomials involved are over $\R$ (making use of the isomorphism between $\C$ and $\R^2$). A function $h:\R^m\rightarrow\R^n$ will be called \emph{algebraic} over a subfield $\F\subseteq\R$ if for each of its $n$ component functions $h_i$ there exist a polynomial $p_i\in \F[y,x_1,\ldots, x_m]$ such that $y=h_i(x)\Leftrightarrow p_i(y,x_1,\ldots,x_m)=0$. $H_d\subseteq\C^{d\times d}$ denotes the space of Hermitian $d\times d$ matrices and $P_d\subseteq H_d$ is the set of positive definite matrices. By $D_d \subseteq P_d$ we denote the set of density matrices that are non-degenerate and have full rank, and by $\mathcal{U}_d$ the set of $d \times d$ unitary matrices.

\section{Relative entropy constrained sets}\label{Sec:RelativeEntropy}


The \emph{relative entropy} between two density operators $\rho,\sigma \in D_d$ is defined as $S(\rho|\!|\sigma)\coloneqq\tr{\rho\ln\rho}-\tr{\rho\ln\sigma}$ whenever ${\rm supp}(\rho)\subseteq{\rm supp}(\sigma)$ and is $+ \infty$ otherwise. 

\begin{theorem}\label{sigma}

For any $c>0$, $d\geq 3$ the set of $d \times d$ positive definite density operators whose relative entropy is equal to $c$ is nowhere semialgebraic. More precisely, the following sets
\begin{align}
    \mathcal{R}_1 & \coloneqq \big\{ \rho \in D_d \;|\; S(\rho |\!| \sigma) = c \big\} \\
    \mathcal{R}_2 & \coloneqq \big\{ \sigma \in D_d \;|\; S(\rho |\!| \sigma) = c \big\} \\
    \mathcal{R}_3 & \coloneqq \big\{(\rho,\sigma) \in D_d \times D_d \;|\; S(\rho|\!|\sigma)=c\big\} 
\end{align}
are nowhere semialgebraic.
\end{theorem}

\begin{proof}
We distinguish between the following cases:
\begin{enumerate}
    \item That for any positive definite density matrix $\sigma \in H_d$ and any open subset $U \subseteq H_d$, the set $\mathcal{R}_1 \coloneqq \big\{ \rho \in D_d \;|\; S(\rho |\!| \sigma) = c \big\} \cap U $ is not semialgebraic in $H_d$ unless it is empty, was established in \cite{Vjosa}.\vspace{7pt}

    \item We now examine the case where the roles of $\rho$ and $\sigma$ are reversed. For any positive definite  density matrix $\rho \in H_d$, assume that the set $\mathcal{R}_2 \coloneqq \big\{ \sigma \in D_d \;|\; S(\rho |\!| \sigma) = c \big\}$ is semialgebraic everywhere. We proceed by contradiction. 
    
    The above set can be rewritten as $\mathcal{R}_2 \coloneqq \big\{ \sigma \in D_d \;|\; \tr{\rho \ln \sigma} = \Tilde{c} \big\}$, where $\Tilde{c} \coloneqq -c - S(\rho)$. Let $V$ denote an open subset in $H_d$. Any $\sigma \in \mathcal{R}_2 \cap V$ can be written as  $\sigma = U \mathrm{diag} (\sigma) U^*$ for some unitary $U \in \mathcal{U}_d$. By Lemma $3$ in \cite{Vjosa} there exists a local algebraic diffeomorphism $\Phi: \sigma \mapsto (D:= \mathrm{diag}(\sigma),U)$ which maps each $\sigma \in \mathcal{R}_2 \cap V$ to a vector whose $d$ first components are the eigenvalues of the density matrix $\sigma$. After specifying a unitary $U$, the set
\begin{equation}
   \mathcal{M} \coloneqq \big\{ \lambda \in \R_{>0}^{d^2} \;|\; \sum_{i=1}^d \lambda_i =1, \sum_{i=1}^d a_i \log \lambda_i = \Tilde{c} \big\}
\end{equation}
where $a_i := (U^* \rho U)_{ii}$, is semialgebraic according to Lemma \ref{lemma:fixedsemi} in the Appendix \ref{AppendixB}. Note that $\mathcal{M}$ is a smooth submanifold of $\R^{d^2}$.
As we will show later, we can always choose $a_1, a_2, a_d \neq 0$ such that $\frac{a_1 + a_d}{a_2} \notin \Q$. For any $\lambda \in \R^{d^2-1}_{>0}$ we define

\begin{align}\label{CHOKO}
    f(\lambda) \coloneqq  a_1 \log \lambda_1 + a_2 \log \lambda_2
    + a_d \log (1-\sum_{i=1}^{d-1} \lambda_i) - \Tilde{c} + \sum_{i=3}^{d-1} a_i \log \lambda_i.
\end{align}
The manifold $\mathcal{M}$ is characterized by the positive roots of the above equation. After fixing $\lambda_3, ..., \lambda_{d-1}$ to $x_3, ..., x_{d-1}$ by a further application of Lemma \ref{lemma:fixedsemi}, the set defined by $f(\lambda_1, \lambda_2, x_3,...,x_{d-2})=0$ defines an algebraic curve. In this situation, the roots of equation \eqref{CHOKO} give rise to the local algebraic curve
\begin{equation}\label{EquBeta}
    \lambda_1^{a_1} \lambda_2^{a_2}(c-\lambda_1 - \lambda_2)^{a_d} = \beta
\end{equation}
for some $\beta, c \in \R$. The implicit function theorem guarantees the existence of the function $\lambda_2=g(\lambda_1)$ as a solution of \eqref{EquBeta}, for a function $g$. 

Moreover, by a standard argument based on analytic continuation, the whole Riemann surface characterized by \eqref{EquBeta} still forms an algebraic curve. We observe that there exists at least one branch of the Riemann surface such that $|\lambda_1| \to \infty$ and $\lambda_2 \sim \lambda_1^{-\frac{a_1+a_d}{a_2}}$ with one particular branch of the complex function $z \mapsto z^{-\frac{a_1+a_d}{a_2}}$. Here and in the following, we write $ f \sim g$ to denote the asymptotic equality of two functions $f,g$ at a point $a \in \C \cup \{ \infty\}$, i.e., $\lim_{z \to a} f(z)/g(z) = w$ for some $w \in \mathbb{C}$.

Coming back to the proof, we consequently find an unbounded open set $U \subset \C^2$ and an algebraic function $g$ such that a pair of complex numbers $(\lambda_1, \lambda_2) \in U$ is a solution to  (\ref{EquBeta}) if and only if we have functional dependence $\lambda_2 = g(\lambda_1)$ and, moreover, the asymptotic equality $g(\lambda_1) \sim \lambda_1^{-\frac{a_1+a_d}{a_2}}$ as $\lambda_1 \to \infty$ holds true in $U$. Considering the inversion $w(x) \coloneqq \frac{1}{g(1/x)}$, we obtain an algebraic function $w(\lambda_1) \sim \lambda_1^{\frac{a_1+a_d}{a_2}}$ as $\lambda_1 \to 0$. This leads to a nonalgebraic singularity at $0$, since we have chosen $\frac{(a_1+a_d)}{a_2} \notin \Q$. This is not possible due to Theorem \ref{lemma:algebraicsingularities} in the Appendix \ref{trivial}.

 It remains to show that in any neighborhood of a unitary $U_0$, there exists another unitary $U$ which satisfies $a_1, a_2, a_d \neq 0$ such that $\frac{a_1 + a_d}{a_2} \notin \Q$ with the notation from above. Note that the submanifold of unitaries characterized by $a_i = 0$
 is a set of measure zero with respect to the Haar measure.   Similarly, for $d \geq 3$ and a fixed $\lambda \in \Q$
 the set of unitaries with $a_1 + a_d = \lambda a_2$ is again a set of measure zero. This is still true for the union over all rational $\lambda \in \Q$. In particular, the set of all unitaries with $a_1, a_2, a_d \neq 0$ such that $\frac{a_1 + a_d}{a_2} \notin \Q$  forms a dense subset, completing the proof.\vspace{7pt}

 \item For the last case, we study the set $\mathcal{R}_3 \coloneqq \big\{(\rho,\sigma) \in D_d \times D_d \;|\; S(\rho|\!|\sigma)=c\big\} \cap V,$ for any open subset $V \subseteq H_d \times H_d.$ Suppose the set $\mathcal{R}_3$ is semialgebraic. By Lemma \ref{lemma:fixedsemi} in the Appendix \ref{AppendixB}, for any $\sigma_0 \in D_d$, the set $\big\{ \rho \in D_d \;|\; S(\rho |\!| \sigma_0)=c \big\}$ would also be semialgebraic. This contradicts $(1)$.
\end{enumerate}
\end{proof}

\begin{remark}

\begin{enumerate}
        \item We have used the natural logarithm for the formulation of the theorem and its proof but note that the same result hold for any other base of the logarithm since the change in the base of the logarithm corresponds to a change in the value of $c$. \vspace{4pt}
        \item In the theorem above, only the case of equality " = c" is considered, but the same result holds for the inequalities " $< c$", " $ \leq c$", "$ > c$" and "$ \geq c$" since the boundaries of semialgebraic sets are semialgebraic. \vspace{4pt}
        \item The same result holds if classical relative entropy is used instead of quantum relative entropy. \vspace{5pt}
    \end{enumerate}
\end{remark}

\begin{figure}[t]
    \centering
    \includegraphics[width=1\textwidth]{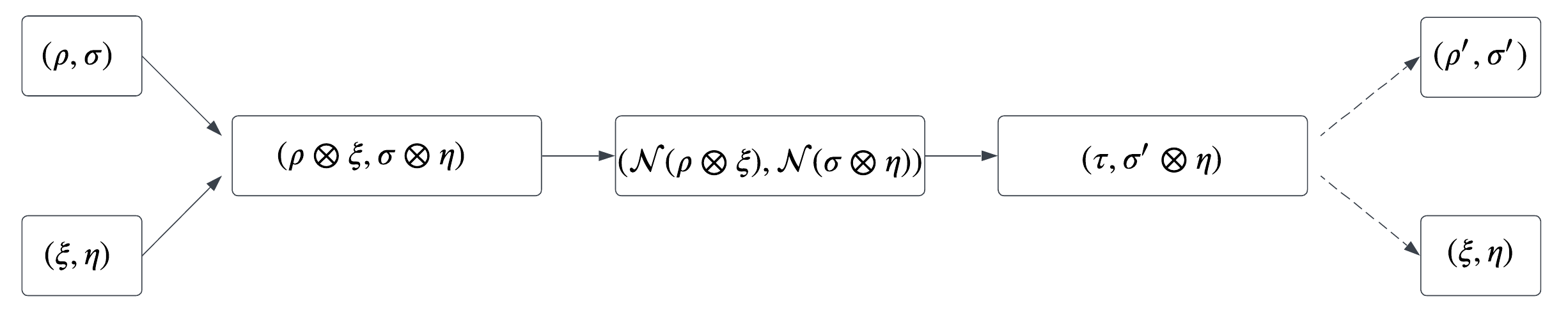} 
    \caption{The pair of commutative quantum states $(\rho,\sigma)$ is transformed into the other pair of commutative quantum states $(\rho',\sigma')$ via the classical channel $\mathcal{N}$ using a classical catalyst $(\xi, \eta)$, where $\eta$ is the uniform distribution on the support of $\xi$. The distribution $\tau$ obtained by applying the channel $\mathcal{N}$ to $\rho \otimes \xi$ has marginals $\rho'$ and $\xi$, respectively. This figure is adjusted from \cite{RelativeEntropy}. } \label{fig:catalytic}
\end{figure}

\emph{Applications:} An immediate application of the above theorem would be to examine the results presented in \cite{RelativeEntropy} from this perspective. There, relative entropy was given an operational meaning beyond its conventional interpretation in the asymptotic framework. It has been shown that the catalytic transformation between pairs of quantum states can be characterized using only the relative entropy. Specifically, given two pairs of commutative quantum states $(\rho,\sigma)$ and $(\rho',\sigma')$, the pair $(\rho, \sigma)$ is transformed to $(\rho', \sigma')$ by using a catalyst consisting of a pair of distributions $(\xi, \eta)$ in conjunction with $(\rho,\sigma)$. The target pair is then generated via a classical channel $\mathcal{N}$ acting on $(\rho \otimes \xi, \sigma \otimes \eta)$ such that the first and second marginals of $\mathcal{N}(\rho \otimes \xi)$ are $\rho'$ and $\xi$, respectively, and $\mathcal{N}(\sigma \otimes \eta) = \sigma' \otimes \eta$, where $\eta$ is the uniform distribution on the support of $\xi$ (see Fig.\ref{fig:catalytic}). 

Moreover, $\mathcal{N}(\rho \otimes \xi)$ is required to be close in relative entropy to the product of its marginals, $\rho' \otimes \xi$. Whether this is also true for pairs of general quantum states and quantum catalysts remains an open question. However, we would like to elaborate on a scenario in which the distinction between the semialgebraic and the transcendental world provides insight into such transformations. Note that the same operational meaning for the relative entropy as above applies if the condition of proximity between $\mathcal{N}(\rho \otimes \xi)$ and $\rho' \otimes \xi$ is now replaced by e.g., the trace distance between $\mathcal{N}(\rho \otimes \xi)$ and $\rho' \otimes \xi$
\begin{equation}
    T(\mathcal{N}(\rho \otimes \xi),\rho' \otimes \xi) \coloneqq \frac{1}{2}|\!| \mathcal{N}(\rho \otimes \xi)- \rho'\otimes \xi   |\!|_1 .
\end{equation}
Indeed, assuming that $\forall \gamma >0$, $\exists$ a channel $\mathcal{N}$ such that $D(\mathcal{N}(\rho \otimes \xi)|\!| \rho' \otimes \xi) < \gamma$, Pinsker's inequality \cite{Pinsker} yields
\begin{equation}
    T(\mathcal{N}(\rho \otimes \xi),\rho' \otimes \xi) \leq \sqrt{ \gamma}.
\end{equation}

Conversely, for any $\beta \in (0,1)$ assume that $T(\mathcal{N}(\rho \otimes \xi),\rho' \otimes \xi) \leq \beta$. Using the continuity bound given by Winter \cite{WinterA} and the Remark $5.10$ from \cite{GCB22} we have:
\begin{equation}
    |D(\mathcal{N}(\rho \otimes \xi)|\!| \rho' \otimes \xi)| \leq \beta \log |A| + \sqrt{2 \beta} \eqqcolon \gamma,
\end{equation}
where $A$ denotes the Hilbert space on which $\rho$ and $\rho'$ live. \vspace{3pt}

Now, for an initial pair of commuting states $(\rho, \sigma)$ on $\C^d \times \C^d$, define $\mathcal{R}_n$ as the set of all pairs of commuting states $(\rho', \sigma')$ on $\C^d \times \C^d$ with the property that for any $\varepsilon \in (0,1)$ and $\gamma \in (0,1)$ there is a pair of probability distributions $(\xi, \eta)$ on $\C^n \times \C^n$ and a classical channel $\mathcal{N}$ on $\C^d \otimes \C^n$ such that the reduced states of $\mathcal{N}(\rho \otimes \xi)$ satisfy $\mathcal{N}(\rho \otimes \xi)_2 = \xi$ and $\frac{1}{2} |\!| \rho' - \mathcal{N}(\rho \otimes \xi)_1 |\!|_1 \leq \varepsilon$. Furthermore, $\mathcal{N}(\sigma \otimes \eta) = \sigma' \otimes \eta$, where $\eta$ is the uniform distribution on the support of $\xi$ and $T(\mathcal{N}(\rho \otimes \xi), \mathcal{N}(\rho \otimes \xi)_2 \otimes \xi) < \gamma$. In other words, $\mathcal{R}_n$ is the set of pairs of states that can be reached (approximately) from $(\rho, \sigma)$ with the help of an $n$-dimensional ‘catalyst’. As a second set, we consider the set $\mathcal{R}$ of all pairs of states $(\rho',\sigma')$ on $\C^d \times \C^d$ whose relative entropy is less than or equal to that of $(\rho, \sigma)$. The question would be whether these two sets are the same for some $n$ as a function of $d$. One way to answer this question is to state that the set $\mathcal{R}_n$ is semialgebraic according to Tarski-Seidenberg, while $\mathcal{R}$ is not, as we saw above. This distinction excludes the equality between $\mathcal{R}_n$ and $\mathcal{R}$ and any kind of semialgebraic characterization of the relative entropy, as long as bounded ancillary systems are considered.

\section{Mutual information constrained sets}\label{Sec:MutualInfo}

To lighten the notation, in this section, we denote by $\mathcal{H}_A$ and $\mathcal{H}_B$ two complex  Hilbert spaces of finite dimension $d_A$ and $d_B$ for systems A and B, respectively. By $\mathcal{S}(\mathcal{H}_{\cdot}) \equiv \mathcal{S}_{\cdot}$ we denote the set of states associated with the Hilbert space $\mathcal{H}_{\cdot}$. The \textit{quantum mutual information} of a bipartite state $\rho_{AB} \in \mathcal{S}(\mathcal{H}_{A} \otimes \mathcal{H}_B)$, quantifying the correlations between subsystems $A$ and $B$, is defined as  
\begin{equation}
    I(\rho_{AB})=I(A:B) \coloneqq S(\rho_A) + S(\rho_B) - S(\rho_{AB}) = S(\rho_{AB}||\rho_A \otimes \rho_B), 
\end{equation}
where $S(\rho_A)$ and $S(\rho_B)$ denote the von Neumann entropy of the marginals $\rho_A \in \mathcal{S}(\mathcal{H}_A)$ and $\rho_B \in \mathcal{S}(\mathcal{H}_B)$ of $\rho_{AB}$, respectively. \vspace{4pt}

Recall that $0 \leq I(A:B) \leq 2 \log[\min \{d_A, d_B\}]$, where the lower bound follows from the positivity of the relative entropy and the upper bound is due to the strong subadditivity of the relative entropy. The set of states that have extremal mutual information $(0$ or $2 \log [\min\{d_A,d_B\}])$ are (semi)algebraic in each dimension $d$. 

Due to the strict positivity of the relative entropy, the case $c=0$ coincides with $\rho_{AB} = \rho_A \otimes \rho_B$. Consequently, this set is characterized by linear constraints and is thus (semi)algebraic. For the other extreme value, we distinguish the cases: $(i) \, d_A = d_B$ and $(ii) \, d_A \neq d_B$. 

For the first case, note that the level sets consist of pure states $\rho_{AB}$ whose marginal $\rho_{A}$ satisfies $S(\rho_A)=\log d_A$. Both these conditions are (semi)algebraic. For the second case, assume without loss of generality that $d_A < d_B$. Note that we can write $I(A:B) = S(A)-S(A|B)_{\rho}=S(A)+S(A|E)_{\psi}$, where $\ket{\psi}_{ABE} \equiv \ket{\psi}$ is a purification of the state $\rho_{AB}$ to some environment $E$ (following the proof of Theorem $11.5.1$, \cite{WildeBook}). We take advantage of the fact that (see Exercise $11.8.9.$ in \cite{WildeBook})
\begin{equation}
 S(A|E) = - S(\Tr_B[\ket{\psi}\bra{\psi}] |\!| (\frac{\1_A}{d_A} \otimes \Tr_{AB}[\ket{\psi}\bra{\psi}])) + \log d_A \end{equation}
is maximal ($\log d_A$) if and only if $\Tr_B[\ket{\psi} \bra{\psi}] = \frac{\1_A}{d_A} \otimes \Tr_{AB}[\ket{\psi} \bra{\psi}]$. Then, the level set of mutual information for  $c = 2 \log d_A$ has the following form:
\begin{align*}
 \mathcal{I} = \big\{ \rho_{AB} \in \mathcal{S}(\mathcal{H}_A \otimes \mathcal{H}_B) \, | \, \exists & \ket{\psi}_{ABE} \, \, \mathrm{s. \, t.} \,  \Tr_E[\ket{\psi}\bra{\psi}]=\rho_{AB}, \, \\
 & \Tr_B[\ket{\psi}\bra{\psi}] = \frac{\1_A}{d_A} \otimes \Tr_{AB}[\ket{\psi}\bra{\psi}] \big\}.
\end{align*}
An application of the Tarski-Seidenberg theorem to the above set shows that it is semialgebraic. For other values of $c$ and other dimensions, the answer is given in the following theorem.

\begin{theorem}
Let $\min(d_A, d_B)\geq 3$ and $c \in (0, 2 \min\{ \log d_A, \log d_B\})$. Then the set of density matrices with constrained mutual information \begin{equation}\label{eq:relentttlevelset}
   \mathcal{I}\coloneqq \big\{\rho_{AB} \in \mathcal{S}(\mathcal{H}_A \otimes \mathcal{H}_B) \;|\; I(\rho_{AB}) \coloneqq S(\rho_{AB}|\!|\rho_A \otimes \rho_B)=c\big\}
\end{equation} where $\rho_A\coloneqq\Tr_B{[\rho_{AB}]}$ and $\rho_B \coloneqq \Tr_A{[\rho_{AB}]}$ is not semialgebraic unless it is empty. 
\end{theorem}
\begin{proof}

 Without loss of generality $ 3 \leq d_A \leq d_B$ and let $0 < c < 2 \log d_A$ be a fixed real number. We proceed by contradiction and assume that the level set of the quantum mutual information is a nonempty semialgebraic set for our choice of $c$.
If the level set $$\mathcal{I}\coloneqq I^{-1}(c) \coloneqq \big\{\rho_{AB} \in \mathcal{S}(\mathcal{H}_A \otimes \mathcal{H}_B) \;|\; I(\rho_{AB})=c\big\}$$ was a semialgebraic set, then so would be ($\S$ 2, \cite{RAG98}) the set 
\begin{align*}
     \mathcal{C} \coloneqq {} & \big\{(\rho_{AB}, \rho_A, \rho_B) \in \mathcal{S}(\mathcal{H}_A \otimes \mathcal{H}_B) \times \mathcal{S}(\mathcal{H}_A) \times \mathcal{S}(\mathcal{H}_B) \;|\; I(\rho_{AB})=c\big\} \\
 & \cap \big\{(\rho_{AB}, \rho_A, \rho_B) \in \mathcal{S}(\mathcal{H}_A \otimes \mathcal{H}_B) \times \mathcal{S}(\mathcal{H}_A) \times \mathcal{S}(\mathcal{H}_B) \;|\; \Tr_B[\rho_{AB}]=\rho_A\big\} \\
 & \cap \big\{(\rho_{AB},\rho_A, \rho_B) \in \mathcal{S}(\mathcal{H}_A \otimes \mathcal{H}_B)\times \mathcal{S}(\mathcal{H}_A) \times \mathcal{S}(\mathcal{H}_B) \;|\; \Tr_A[\rho_{AB}]=\rho_B\big\}.
 \end{align*}
\vspace{5pt}
If we further intersect $\mathcal{C}$ with the semialgebraic set 
$$\mathcal{P} \coloneqq \big\{ \rho_{AB} \in \mathcal{S}(\mathcal{H}_A \otimes \mathcal{H}_B) \;|\; \tr{\rho_{AB}^2} = 1 \big\},$$ we get another semialgebraic set. Now, for any $( \rho_{AB}, \rho_A, \rho_B) \in \mathcal{M} \coloneqq \mathcal{C} \cap \mathcal{P}$ we have $I(\rho_{AB}) = 2 S(\rho_A)$, and therefore $\mathcal{M}$ takes the form
\begin{align*}
    \mathcal{M} = \big\{ (\rho_{AB}, \rho_{A}, \rho_{B}) \in  \mathcal{S}(\mathcal{H}_A \otimes \mathcal{H}_B) \times \mathcal{S}(\mathcal{H}_A) \times \mathcal{S}(\mathcal{H}_B) & \;|\; \\ 
    \Tr_B[\rho_{AB}]  = \rho_A,  \Tr_A[\rho_{AB}] = \rho_B, \tr{\rho_{AB}^2}=1, S(\rho_A) = \frac{c}{2}  \big\}.
\end{align*}
As a corollary of the second form of the Tarski-Seidenberg theorem (2.1.2, \cite{IntroSemiAlgGeo}) the image of $\mathcal{M}$ by the projection on the space of the second coordinate, which is given by $\big\{ \rho_A \in \mathcal{S}(\mathcal{H}_A) \;|\; S(\rho_A) = \Tilde{c} \big\}$ due to the well-known purification lemma, where $\Tilde{c}\coloneqq\frac{c}{2}$, is a semialgebraic set. This is impossible since the von Neumann entropy-constrained sets for dimensions $d \geq 3$ are transcendental everywhere \cite{Vjosa}. 
\end{proof}

\begin{remark}
The same result is valid if classical mutual information is used instead of quantum mutual information.
\end{remark}


\section{R\'{e}nyi entropy constrained sets}\label{sec:Renyi}

For $\rho \in D_d $ the quantum R\'{e}nyi entropy of order $\alpha \in (0,1) \cup (1, \infty)$ is defined as
\begin{equation}
    S_{\alpha}(\rho) \coloneqq \frac{1}{1-\alpha} \log_b \tr{\rho^{\alpha}}.
\end{equation}

As generally known, in the limit $\alpha \to 1$ the R\'{e}nyi entropy reduces to the von Neumann entropy, the level set of which was fully analyzed in \cite{Vjosa}. We will extend these results to the $\alpha-$R\'{e}nyi entropy-constrained sets for any $\alpha \geq 0$. We start with the two remaining limit cases.\vspace{2pt}

\begin{enumerate}
    \item \textit{Max entropy (Hartley entropy)}:  Let $\mathcal{S}(\mathcal{H})$ denote the set of density operators for the Hilbert space $\mathcal{H}$. Then, the max entropy for $\rho \in \mathcal{S}(\mathcal{H})$ is defined as follows
\begin{equation}
    S_0(\rho) \coloneqq \lim_{\alpha \to 0} S_{\alpha}(\rho) = \log_b \mathrm{rank}(\rho).
\end{equation}    
For any $a \in \R$ and $d \geq 2$ the level set $\mathcal{S}_0 \coloneqq \{ \rho \in \mathcal{S}(\mathcal{H}) \;|\; S_0(\rho) = a \} = \big\{ \rho \in \mathcal{S}(\mathcal{H}) \;|\; \mathrm{rank}(\rho) = c \big\}$, where $c \coloneqq b^a \in \R$, is semialgebraic everywhere because the rank of the density matrix $\rho$ is the size of the largest non-vanishing minor \cite{WolfCubbitPerez}. \vspace{4pt}
    
    \item \textit{Min entropy}: 
\begin{equation}
    S_{\infty}(\rho) \coloneqq \lim_{\alpha \to \infty} S_{\alpha}(\rho) = \log_b |\!| \rho |\!|,
\end{equation}    
  
where $|\!| \cdot |\!|$ denotes the operator norm. For any $a \in \R$ and $d \geq 2$, the set $\mathcal{S} \cap W \coloneqq \big\{ \rho \in D_d \;|\;  |\!| \rho |\!| = b^a \big\} \cap W$ is again semialgebraic for any open set $W \subset H_d$, as the set of Hermitian matrices with bounded norm \cite{WolfCubbitPerez}.
\end{enumerate}\vspace{5pt}

Let us further remark that for any $\alpha \in (0,1) \cup (1, \infty)$ the set of states with extremal R\'{e}nyi entropy $0$ or $\log d$ are (semi)algebraic in each dimension $d$. Indeed, the states with vanishing R\'{e}nyi entropy are exactly the pure states and $S_{\alpha}(\rho) = \log d$ corresponds to the set of the maximally mixed states, both of which exhibit an algebraic characterization. Similarly, for $d =2$ all level sets are semialgebraic since any constraint of the form $S_{\alpha}(\rho) = c$ is equivalent to a constraint on the eigenvalues $\{\lambda_1,\lambda_2 \}= \{\gamma, 1- \gamma \} $ of $\rho$, which in turn can be formulated as roots of a quadratic polynomial. The following theorem characterizes the remaining nontrivial situations.

\begin{theorem}
For any $d \geq 3$ and $c \in (0,\log d)$, the set of $d \times d$ density operators with the R\'{e}nyi entropy being fixed to $c$ is transcendental everywhere if  the order $\alpha \notin \Q$, and otherwise it is semialgebraic. That is, if
\begin{equation}\label{eq:relenttttlevelset}
   \mathcal{S}_{\alpha} \coloneqq \big\{\rho \in D_d \;|\; S_{\alpha}(\rho)=c\big\},
\end{equation} 
then for any open subset $W \subset H_d$ the set $\mathcal{S}_{\alpha} \cap W$ is not semialgebraic for $\alpha \in (\R \setminus \Q) \cap [(0,1) \cup (1, \infty)]$ unless it is empty. 
\end{theorem}
\begin{proof}
We distinguish the following cases:
\begin{enumerate}
    \item For $d \geq 3$ and  $\alpha \in \N \cap [(0,1) \cup (1, \infty)]$ we write $S_{\alpha}(\rho) = c$ as $\tr{\rho^{\alpha}}= v$ for $v \coloneqq b^{c(1-\alpha)} \in \R$, which is a polynomial over the field of real numbers. As a consequence, the level sets of the $\alpha-natural$ R\'{e}nyi entropy $\mathcal{S}_{\alpha}$ are everywhere (semi)algebraic. \vspace{5pt}
    
    \item For $d \geq 3$, $\alpha \in \Q \cap [(0,1) \cup (1, \infty)]$ and $a,b \in \N$, the set $\mathcal{S}_{\alpha}=\big\{ \rho \in D_d \;|\; \tr{\rho^{a/b}} = v \big\}$ can be rewritten as 
    \begin{equation}
        \mathcal{S}_{\alpha}\coloneqq \{\ \rho \in D_d \;|\; \exists X \geq 0, X^b=\rho, \tr{X^a} = v \}\
    \end{equation} 
    An application of the Tarski-Seidenberg theorem to the above set yields the claim. \vspace{5pt}

    \item For $d \geq 3$ and  $\alpha \in (\R \setminus \Q) \cap [(0,1) \cup (1, \infty)]$ the set \eqref{eq:relenttttlevelset} reduces to $\mathcal{S}_{\alpha} \coloneqq \big\{ \rho \in D_d \;|\; \tr{\rho^{\alpha}} = v \big\}$, $v \coloneqq b^{c(1-\alpha)}$, which we assume to be everywhere semialgebraic. The proof then follows the same idea as that of relative entropy. Everything boils down to the analysis of the implicit equation
    \begin{equation}\label{eq:irrational}
        \lambda_1^{\alpha} + \lambda_2^{\alpha} + (\gamma-\lambda_1 - \lambda_2)^{\alpha} = \beta
    \end{equation}
    which by assumption defines an algebraic curve, where $\gamma \coloneqq 1 - \sum_{i=3}^{d-1} \lambda_i \in \R$, and $\beta \coloneqq v - \sum_{i=3}^{d-1 } \lambda_i \in \R$. Note that we can assume $\beta \neq 2 (\gamma/2)^{\alpha}$ in the case $c \notin \{0, \log d \}.$
    
    We claim that there is a complex solution of equation (\ref{eq:irrational}) with $\lambda_1 + \lambda_2 = \gamma$, which we denote by $(x,y \coloneqq \gamma -x)$. This follows from the following observations. Considering the function $f\colon \C \to \C$
    defined by $f(z) \coloneqq z^{\alpha} + (\gamma - z)^{\alpha}$ -- or to be more precise its branches -- the open mapping theorem yields that its range is an open subset of the complex plane $\C$. On the other hand, the irrationality of $\alpha$ and the resulting infinite branches imply the density of its range. Indeed, let us write $z = |z| e^{i \phi}$ and $z' = \gamma - |z| = |z'| e^{i \phi'}$ and we observe that if $\phi$ and $\phi'$ are irrational and rationally independent, all branches of $f$ together lead to an image which is dense in the annulus $\{ w \, | \,\, \, | \, |z|^{\alpha} - |z'|^{\alpha}| \leq |w| \leq |z|^{\alpha} + |z'|^{\alpha} \}.$ Making use of the transcendence of the trigonometric functions, such complex numbers $z$ are themselves dense in $\C$. Combining these considerations, one concludes the existence of a solution $(x,y)$ as above.

    The idea is now that if we fix a branch of the algebraic curve \eqref{eq:irrational} which contains the solution $(x,y)$, then we find a non-algebraic behavior close to $(x,y)$ which amounts to the desired contradiction. Let us make this intuition precise. First, the tuple $(x,y)$ is regular, as the partial derivatives do not vanish. Thus, in a sufficiently small neighborhood of $(x,y)$ the solutions of \eqref{eq:irrational} 
    may be represented in the form $y' = g(x')$, where $g$ is by assumption an algebraic function. We now expand the function $g$ around $x$ using its characterizing identity \eqref{eq:irrational}. We proceed iteratively until we "detect" the non-algebraic singularity. Let us demonstrate the first step. Note that $x \neq y$ and we assume that both $x,y$ do not vanish. A first-order Taylor expansion yields 
    \begin{equation}\label{eq:tay1} 
    \begin{split}
        &\alpha x^{\alpha-1} (x-x') + \alpha y^{\alpha-1} (y-y')  + (x+y-x'-y')^{\alpha} \\
        &= o(x-x') + o(y-y').
    \end{split}
    \end{equation} 
   If $ \alpha  < 1$, one  sees that the left-hand side of \eqref{eq:tay1} can only cancel up to the first order if $(y-y') \sim (x-x')^{\alpha}$, which shows that $g$ cannot be an algebraic function. If $\alpha > 1$, one obtains $(y-y') = -\frac{x^{\alpha-1}}{y^{\alpha-1}} (x-x') + o(x-x')$, which we plug into \eqref{eq:irrational} and continue with a second order Taylor expansion. 
   After finitely many steps,  we arrive at an expansion of the form 
    $$ (y'-y) \sim \sum_{k} c_k (x'-x)^k + \delta (x'-x)^{\alpha} + o((x-x')^{\alpha}) $$ 
    for some complex constants $c_k$ and $\delta \neq 0$. It easily follows then that the $(k+1)th$ derivative of $g$ has a non-algebraic singularity at $x$. Theorem \ref{lemma:algebraicsingularities} from the Appendix \ref{trivial} completes the proof. 
     
    \end{enumerate}
\end{proof}
\begin{remark}
From the above proof, it is clear that the same result holds if classical R\'{e}nyi entropy is used instead of quantum R\'{e}nyi entropy.    
\end{remark}

\section{Outlook}
Variants (or in some cases direct consequences) of the arguments presented here and in \cite{Vjosa} can be extended to divergence measures that appear in classical and quantum information theory.   \vspace{8pt}


\emph{Acknowledgments:} The authors thank Michael M. Wolf for insightful discussions and his feedback on a draft of this paper, as well as  \'{A}lvaro M. Alhambra, Cambyse Rouz\'{e}, Paul Gondolf, and Zahra Baghali Khanian for their helpful input. VB acknowledges support from the International Max Planck Research School for Quantum Science and Technology at the Max-Planck Institute of Quantum Optics. CM acknowledges support from the Deutsche Forschungsgemeinschaft (DFG, German Research Foundation) under Germany’s Excellence Strategy – EXC-2111 – 390814868.


\appendix

\section{Algebraic functions}\label{trivial}
In this section, we summarize some characterizations of algebraic functions that we use to show the transcendental nature of the entropy-constrained sets studied in this paper. One of the most well-known characterizations of algebraic functions is that they have a compact Riemann surface \cite{AlgebraicCurves}.

Another characterization of the algebraic functions is given in terms of \textit{algebraic singularities} \cite{NonAlgebraicFunction}.

\begin{definition}(\cite{NonAlgebraicFunction})
A singular point $z_0$ of $f(z)$ is called \textit{algebraic} if in a neighborhood of $z_0$ the function can be represented by a Puiseaux series
$$f(z)= \sum_{l=M}^{\infty}a_l (z-z_0)^{l/n}$$
where $n (n>0)$ and $M$ are integers, and $a_M \neq 0$. When $M < 0$, this point is called a \textit{critical pole}.
\end{definition}
To show the transcendence of $\alpha$-R\'{e}nyi and relative entropy-constrained sets, we use the following theorem and look for the contradiction to the fact that non-algebraic functions have more than algebraic singularities.

\begin{theorem}(Theorem 3.1, \cite{NonAlgebraicFunction})\label{lemma:algebraicsingularities}
A global analytic function $f(z)$, $z \in \C \cup \{ \infty \}$ is algebraic if and only if all its singular points are isolated and it has finitely many algebraic singular elements.
\end{theorem}

\section{Semialgebraic sets}\label{AppendixB}
  
Let $\Psi(X,Y)$ be a first-order formula (2.1.2, \cite{IntroSemiAlgGeo}). The following holds:

\begin{lemma}\label{lemma:fixedsemi}
$A \coloneqq \{ (\underline{x},\underline{y}) \in \R^{m+n} \;|\; \Psi(\underline{x},\underline{y}) \}$ semialgebraic $\Longrightarrow$ $\forall \underline{x_0} \in \R^m$ $\{ \underline{y} \in \R^n \;|\; (\underline{x_0},\underline{y}) \in A\}$ is semialgebraic.
\end{lemma}

\begin{proof}
Follows directly from the definition of semialgebraic sets.
\end{proof}

\begin{remark}
The same result holds if we fix any other component from the entries of the defining set.
\end{remark}

\vspace{10pt}

\bibliographystyle{ieeetr}
\bibliography{Bakery}{}

\end{document}